\documentclass[pra,twocolumn,superscriptaddress,leqn]{revtex4-2}
\usepackage{amsmath,amssymb,amsfonts,bbm,graphicx,times,psfrag}
\usepackage[pdftex]{color}
\usepackage{amsmath,bm}
\usepackage[colorlinks, linkcolor=blue, citecolor=blue, urlcolor=blue, breaklinks=true]{hyperref}
\pdfoutput=1 
\usepackage{microtype}

\usepackage{csquotes}
\usepackage{color}

\usepackage{physics}

\begin{document}

\title {Amplifying Two-Mode Squeezing in Nanomechanical Resonators} 
	
	\author{Muhdin Abdo Wodedo}
	\affiliation{Department of Applied Physics, Adama Science and Technology University, 1888, Adama, Ethiopia.}
	\author{Tesfay Gebremariam Tesfahannes}%
	\email{tesfaye.gebremariam@amu.edu.et}
	\affiliation{Department of Physics, Arba Minch University, 21, Arba Minch, Ethiopia.}
	\author{Tewodros Yirgashewa Darge}
	\affiliation{Department of Applied Physics, Adama Science and Technology University, 1888, Adama, Ethiopia.}
\author{Mauro Pereira}
\affiliation{College of Engineering and Physical Sciences, Department of Physics, Khalifa University of Science and Technology, 127788, Abu Dhabi, UAE}
   \author{Berihu Teklu}
 \email{berihu.gebrehiwot@ku.ac.ae}
	\affiliation{College of Computing and Mathematical Sciences, Khalifa University, 127788, Abu Dhabi, United Arab Emirates.} 
\affiliation{Center for Cyber-Physical Systems (C2PS), Khalifa University, Abu Dhabi 127788, United Arab Emirates.}

 \date{\today}
 
\begin{abstract}
Quantum squeezing plays a crucial role in enhancing the precision of quantum metrology and improving the efficiency of quantum information processing protocols. We thus propose a scheme to amplify two-mode squeezing in nanomechanical resonators, harnessing parametric amplification and two-tone laser controls. The red-detuned laser drives facilitate the cooling of the nanomechanical resonators down to their ground state and allow optimal quantum state transfer in the weak-coupling, resolved sideband regime. In particular, the competing blue-detuned lasers in the driving pairs induce displacement squeezing in mechanical resonators. Thus, the quantum state transfer of the squeezing in nanomechanical resonators and the intracavity correlated photons of the parametric amplifier significantly enhance the two-mode mechanical squeezing. Notably, increasing the coupling strength of the red detuned laser and the ratio of blue-to-red detuned laser dramatically amplifies the two-mode mechanical squeezing under realistic experiment parameters of a typical optomechanical system. Our findings reveal that the proposed cooperative mechanism effectively enhances the level of two-mode mechanical squeezing with a considerable improvement and demonstrates exceptional resilience to thermal noise.
\end{abstract}

\maketitle    

\section{Introduction}

Quantum metrology enhances the sensitivity of measurement beyond the classical methods. Squeezing further improves this by reducing a quadrature below the standard quantum limit set by shot noise~\cite{b1}. Direct noise reduction is invaluable in quantum metrology, quantum information processing, and quantum computing~\cite{b2,b3,b4,b5}. Therefore, leveraging cavity optomechanical systems to enhance squeezing is timely and valuable. The underlying mechanism in an optomechanical system is that light can precisely control and manipulate mechanical states through radiation pressure, stimulating various dynamic behaviors of interaction with matter~\cite{b6,b7}. Among the numerous research fields that cavity optomechanics can offer, beyond squeezing~\cite{ b8,b9,b10} are e.g., quantum entanglement~\cite{b11,b12,b13,b14,b15,b16}, quantum coherence~\cite{b17}, nonlinearity and nonclassicality in a nanomechanical resonator~\cite{b18}, optomechanically induced transparency~\cite{b19,b20}, and cooling of mechanical resonators~\cite{b21}.

In particular, cavity optomechanics has made great strides in squeezed-state generation in recent years. Scholars have theoretically and practically investigated and presented several techniques for squeezing mechanical oscillators under diverse driving circumstances~\cite{b22, b23, b24}. Thus, parametric pumping~\cite{b25} constrained by system stability or periodic modulation of the external driving amplitude results in weak mechanical squeezing~\cite{b26, b27} that does not exceed the conventional 3 dB threshold have been investigated. Moreover, various approaches have been utilized to squeeze in mechanical resonators, surpassing the limitation. These include two-tone driving~\cite{b28}, {Duffing} nonlinearity~\cite{b29}, strong harmonic mechanical driving~\cite{b30}, broadband squeezed quantum field~\cite{b31}, and others. It is also possible to break the conventional limit through the joint effects of {Duffing} nonlinearity with two-tone drivings~\cite{b32}, intracavity squeezing~\cite{b33}, and two-tone driving fields with mechanical parametric amplification~\cite{b34}. Furthermore, other investigations surpass the 3 dB limit due to broadband squeezed vacuum with parametric pumping~\cite{b35}, two atomic ensembles with two-tone laser fields~\cite{b36}, and an amplitude-modulated laser field with auxiliary cavity~\cite{b37}.

The quantum squeezing engineered by two-tone driving~\cite{b38} was substantiated through experiment. This result stipulated that the mechanical squeezing is enhanced above the 3 dB limit by increasing the blue-to-red-detuned laser's coupling strength ratio to its optimal value. Most recently, the generation of single-mode mechanical squeezing can be achieved through the joint effect of two-tone driving with parametric pumping~\cite{b39} and the Sagnac effect~\cite{b40}. The results indicated that the degree of optimal mechanical squeezing is very far above the 3 dB limit. 

{Recently, essential methods have been implemented to prepare two mechanical oscillators in a two-mode squeezed and entangled state in an optomechanical system to point out them as a resource of quantum information processing. The two-mode mechanical squeezing can be generated through engineering of a single reservoir~\cite{b41}, mechanical modulations~\cite{b42}, and frequency modulation~\cite{b43}. In addition to these, coupling two-mode squeezed state light along with the injection of three-level cascaded atoms into a doubly resonant optomechanical cavity system results in the transfer of quantum features and generates two-mode mechanical squeezing~\cite{b44}. Moreover, two-mode mechanical squeezing enables mechanical entanglement in macroscopic mechanical mirrors with continuous measurement and feedback control~\cite{b45}, and through pulsed optomechanical measurements~\cite{b46}. Generation of two-mode squeezing and entanglement of two microwave fields in a compound four-mode electro-optomechanical system has been addressed~\cite{b47}.}

In light of this, we assert that an additional nondegenerate optical parametric amplifier (NDOPA) can surpass the 3 dB limit, further enhancing a two-mode mechanical squeezing caused by driving each optomechanical cavity with a two-tone laser. However, a two-tone drive to one of the optomechanical cavities in the system cannot beat the conventional limit in two-mode mechanical squeezing. This work proposes a novel scheme that improves two-mode squeezing in nanomechanical resonators by exploiting the effects of two pairs of two-tone driving and nondegenerate parametric pumping. We considered an optomechanical system consisting of NDOPA in a driven, doubly resonant cavity coupled with two mechanical oscillators through radiation pressure. Specifically, we assumed the system works in the weak-coupling and resolved sideband regime. The effect of red-detuned laser drives can cool mechanical resonators down to their ground states, which paves the way for generating squeezed states in mechanical elements. Furthermore, the red-detuned lasers allow quantum state transfer; meanwhile, the competing blue-detuned laser drives induce mechanical displacement squeezing. Thus, the quantum state transfer of each mechanical resonator and the intracavity squeezing of NDOPA significantly enhance the two-mode squeezing in nanomechanical resonators. 

We show that the parametric phase and parametric coupling coefficient of NDOPA, the power of the blue-detuned laser drives, the coupling strength of red-detuned laser drives, and the ratio of coupling strength of blue-to-red detuned laser drives significantly influence the two-mode mechanical squeezing. Moreover, the cooperative effect of two pairs of two-tone driving and parametric pumping contributes to overcoming the conventional 3 dB level for a lower amplitude ratio of blue-to-red detuning laser drives. However, the two pairs of two-tone driving alone cannot beat this conventional limit in the aforementioned parametric regime. Furthermore, increasing the temperature of mechanical baths and the damping rates negatively affects {the two-mode squeezing state of the bipartite mechanical modes.} Thus, the system under our investigation enables a platform for quantum sensing and information processing.

The rest of the paper is organized as follows. In Section~\ref{sec:Model and its dynamics}, the model Hamiltonian and its dynamics, along with the construction of the steady-state covariance matrix, are elaborated. Section~\ref{sec:Quantifying squeezing} describes the method of quantifying squeezing from covariance matrix entries, including the results and discussions. {Moreover, the experimental feasibility of the scheme is indicated in Section~\ref{sec:Experimental feasibility}.} The concluding remarks are given in Section~\ref{sec:Conclusions}. {The supplementary material that completes the main text is referred to \textcolor{blue}{Appendix} part.}

\section{ Model and its dynamics }
\label{sec:Model and its dynamics}

{Recent advances have demonstrated the potential of non-linear parametric amplification to enhance quantum effects in optomechanical systems. In particular, a previously implemented setup that involves single-mode parametric amplification has facilitated the investigation of phenomena such as optical bistability, cavity mechanical entanglement, and single-mode mechanical squeezing~\cite{b48,b49}. Furthermore, it is possible to prepare the mechanical mode into a squeezed steady state by employing reservoir engineering techniques, specifically, two-tone driving of an optical or microwave cavity field coupled to a mechanical resonator. Notably, this approach eliminates the need for fast measurement and feedback control, which are typically required in conventional feedback-based schemes~\cite{b38}.} 
\begin{figure}
    \centering
    \includegraphics[width=\linewidth]{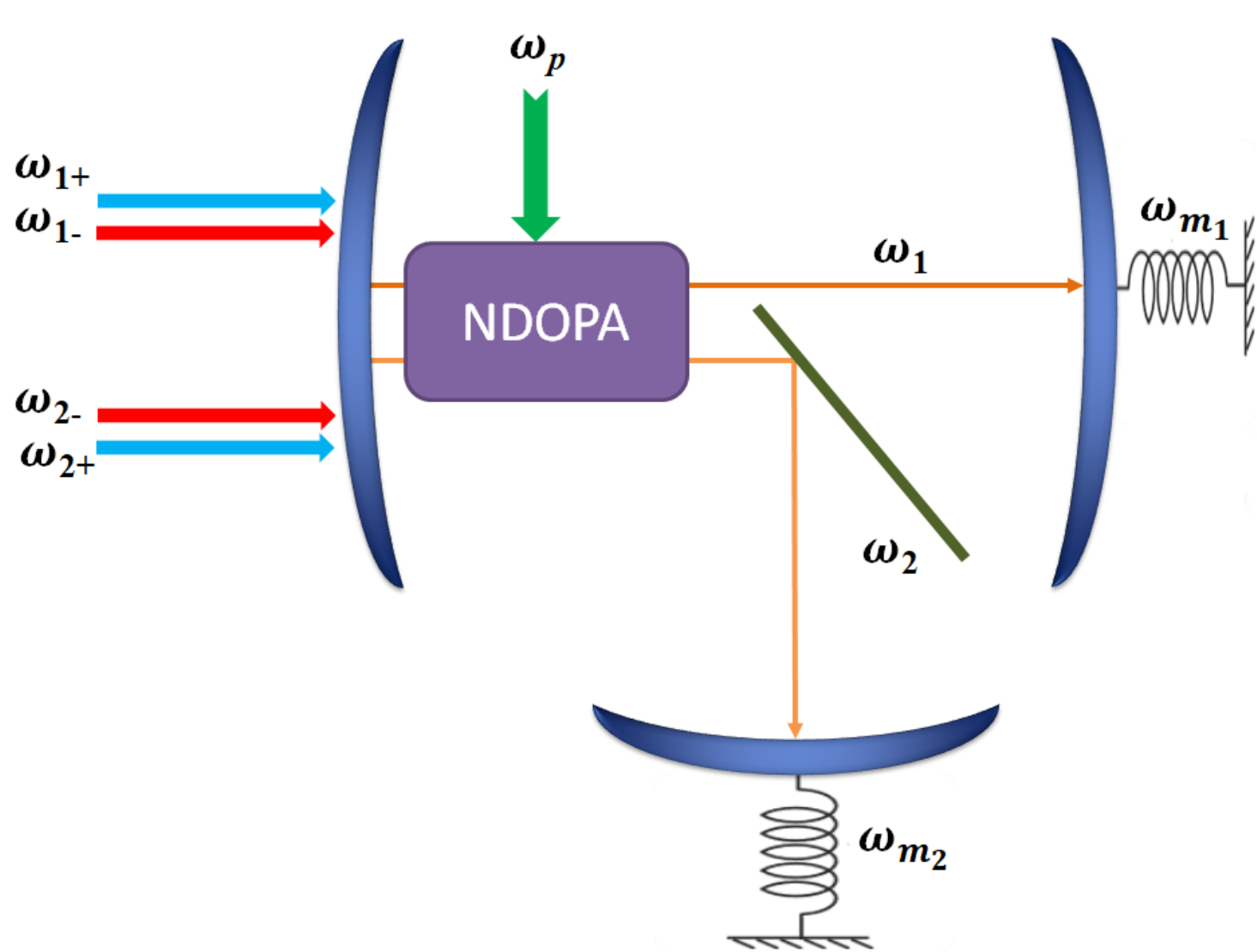}
    \caption{The schematic model of the system, a doubly resonant hybrid optomechanical cavity, comprises a pumped {NDOPA} and two pairs of two-tone laser drives. Further details are provided in the main text}
    \label{fig:1}
\end{figure}

{Specifically, Fig.~\ref{fig:1} illustrates the schematic model of the system under investigation, which consists of NDOPA with second-order nonlinearity embedded in a doubly resonant optical cavity. The cavity is coupled to two perfectly reflecting movable mirrors with effective masses $m_1$ and $m_2$, each oscillating at frequency $\omega_{m_j}$, and one fixed mirror. Each cavity mode has a resonant frequency $\omega_j$ and a corresponding effective cavity length $L_j$. Moreover, each cavity is driven by a pair of laser tones at frequencies $\omega_{j_-} = \omega_j - \omega_{m_j}$ and $\omega_{j_+} = \omega_j + \omega_{m_j}$, corresponding to the red-detuned (anti-Stokes) and blue-detuned (Stokes) sidebands, respectively. The NDOPA is pumped at a frequency $\omega_p$, generating strongly correlated signal and idler photons via down-conversion. These subharmonic photons are nearly resonant with the cavity mode frequencies $\omega_1$ and $\omega_2$. The total Hamiltonian of the present optomechanical system (in units of $\hbar$) is given as}~\cite{b7, b28}
\begin{align}\label{eq:1}
\hat{H}=\sum ^{2}_{j=1}\bigg(\omega_{j} \hat{c}^\dagger_{j}\hat{c}_{j}+\omega_{m_j}\hat{d}^\dagger_{j}\hat{d}_{j}-  g_{j}\hat{c}^\dagger_{j}\hat{c}_{j}(\hat{d}^\dagger_{j}+\hat{d}_{j})&\nonumber\\+\big[ ( E_{j_+} e^{-i\omega_{j_{+}}t}+ E_{j_-} e^{-i\omega_{j_{-}}t })\hat{c} ^\dagger_{j}+H.c\big]\bigg)&\nonumber\\+i \Lambda \big(e^{i\phi } \hat{c}^\dagger_{1}\hat{c}^\dagger_{2} e^{-i\omega_{p}t}-e^{-i\phi}\hat{c}_{1}\hat{c}_{2}  e^{i\omega_{p}t }\big).
\end{align}
Here, the first two terms in the sum correspond to the free Hamiltonian of the cavity and mechanical modes, respectively. The operators $\hat{c}_{j}(\hat{d}_{j})$ and $\hat{c}^\dagger_{j}(\hat{d}^\dagger_{j})$ indicate the annihilation and creation operators of {each optical cavity (mechanical mode)} and obeys the usual commutation relations, $[\hat{O}_{j},\hat{O}_{j'}]=[\hat{O}^\dagger_{j},\hat{O}^\dagger_{j'}]=0, [\hat{O}_{j},\hat{O}^\dagger_{j'}]=\delta_{jj'}$ with $ O= c, d$ and $j,j'=1,2$. The third term in the sum is the Hamiltonian of optomechanical interaction due to radiation pressure and $g_{j} =(\omega_{j}/L_{j}) x_{zpf_j}$ is the single-photon coupling rate with the zero point fluctuation $x_{zpf_j}=\sqrt{\hbar/(2m_{j}\omega_{m_j }})$. The fourth term of the sum represents a {Hamiltonian that emerges} as a result of the laser drives of field strength $|E_{j_{\pm}}|=\sqrt{(\kappa_{j}P_{j_{\pm}})/(\hbar \omega_{j_{\pm}})}$ with $P_{j_{\pm}}$ is laser driving power and $\kappa_{j}$ is the cavity decay rate. The last term of~\eqref{eq:1} refers to the interaction Hamiltonian of the parametric amplifier, where $\Lambda$ is the parametric coupling coefficient proportional to the amplitude of the pumping field at the parametric phase $\phi$. 

The system dynamics, including dissipation, {are governed by} the quantum Langevin equations (QLEs)~\cite{b1}, and can be written as 
\begin{align}\label{eq:2}		
\begin{split} 
\frac{d}{dt}\hat{c}_{j}=&-\big(\frac{\kappa_{j}}{2}+i\omega_{j}\big)\hat{c}_{j}+ig_{j}\hat{c}_{j}(\hat{d}^\dagger_{j}+\hat{d}_{j}) + \sqrt{\kappa_{j}}\hat{c}^{in}_{j}\\&-i(E_{j_+}e^{-i\omega_{j_+}t}+E_{j_-}e^{-i\omega_{j_-}t} )\\&+ \Lambda e^{i\phi }e^{-i\omega_{p}t}\hat{c}^\dagger_{k},\\
\frac{d}{dt}\hat{d}_{j}=&	-\big(\frac{\gamma_{j}}{2}+i\omega_{m_j}\big)\hat{d}_{j}+   ig_{j}\hat{c}^\dagger_{j}\hat{c}_{j} + \sqrt{\gamma_{j}}\hat{d}^{in}_{j},
\end{split}
\end{align}
where $\gamma_{j}$ is the damping rate of the $j^{\textrm{th}}$ mechanical resonator (moving mirror), and $\hat{c}^{\mathrm{in}}_{j}$ and $\hat{d}^{\mathrm{in}}_{j}$ denote the input vacuum and thermal noise operators associated with the optical and mechanical modes, respectively. The indices $j,k=1,2$ with $j \ne k$. For a mechanical resonator of high-quality factor $(Q_{m_j}=\omega_{m_j}/ \gamma_{j}\gg 1)$, and at sufficiently low temperatures $T_{j}$, the mechanical noise {can be Markovian}~\cite{b50}. Thus, the noise operators $\hat{d}^{in}_{j}$ and $\hat{c}^{in}_{j}$ of zero mean with their nonzero delta correlations are  {$\big<\hat{d}^{in,\dagger}_{j}(t)\hat{d}^{in}_{j} (t')\big>=n_{m_j}\delta(t-t')$, $\big<\hat{d}^{in}_{j}(t)\hat{d}^{in,\dagger}_{j}(t')\big>=(n_{m_j}+1)\delta(t-t')$}, $\big<\hat{c}^{in}_{j}(t)\hat{c}^{in,\dagger}_{j}(t')\big>=(n_{c_j}+1)\delta(t-t')$ and $\big<\hat{c}^{in,\dagger}_{j}(t)\hat{c}^{in}_{j}(t')\big>=n_{c_j}\delta(t-t')$ \cite{b51}, where $n_{c_j(m_j)}=\big\{\exp{[\hbar\omega_{c_j(m_j)} /(k_B T_{j})\big]-1 }\}^{-1}$ is the mean thermal bath occupations number of the $j^{th}$ cavity(mechanical) mode, and $k_{B}$ denotes the standard Boltzmann constant. 

For a steady-state mean value of each bosonic operator $O^{s}_{j}$ that is significantly larger than the corresponding quantum fluctuation of the operators $\delta\hat{O}_{j}$, that is $|O^{s}_{j}|\gg \delta\hat{O}_{j}$. Thus, we can adopt a linearized description of the dynamics~\cite{b52} for a zero mean of quantum fluctuation operator around the steady-state value $\big<\delta\hat{O}_{j}\big>=0$, such that $\hat{O}_{j} = O^{s}_{j}+\delta\hat{O}_{j}$. Then, we introduce the mean of cavity field amplitude as $c^{s}_{j}=c^{s}_{j_+}\exp({-i\omega_{j_+}t})+c^{s}_{j_-}\exp({-i\omega_{j_-}t})$ \cite{b28}, with $c^{s}_{j_+}$ and $c^{s}_{j_-}$ are the steady state amplitude of each cavity mode in the long time limit in~\eqref{eq:2} and we have 
\begin{equation}\label{eq:3}	
c^{s}_{j_\pm}=\frac{ iE_{j_\pm}[\frac{\kappa_{k}}{2}-i\Delta_{k_\pm}]}{[\frac{\kappa_{j}}{2}+i\Delta_{j_\pm}][\frac{\kappa_{k}}{2}-i\Delta_{k_\pm}]-\Lambda^2}.
\end{equation}
Here, the high-frequency terms can be omitted reasonably, the expression $\Delta_{j_\pm}=\tilde{\omega}_{j}-\omega_{j_\pm}$ is the effective detuning of each cavity to the laser drives with $\tilde{\omega}_{j}=\omega_{j}-g_{j}(d^{s*}_{j}+d^{s}_{j})$ being the shifted cavity frequency by radiation pressure (optomechanical coupling). In addition, for high mechanical quality factor at steady state, we have $d^{s}_{j}=g_{j}|c^{s}_{j}|^2/\omega_{m_j}$. 

Moreover, the QLEs in~\eqref{eq:2} can be safely linearized by neglecting {the second-order} quantum fluctuation terms, {such as} $\delta\hat{c}^\dagger_{j}\delta\hat{c}_{j}$ and $\delta\hat{c}_{j}(\delta\hat{d}^\dagger_{j}+\delta\hat{d}_{j})$. In this regard, by employing the effective coupling strength, $G_{j}=G_{j_+}\exp({-i\omega_{j_+}t}) + G_{j_-}\exp({-i\omega_{j_-}t})$ and and introducing the slowly varying operators $\delta\tilde{c}_{j}=\delta\hat{c}_{j}\exp({i\tilde{\omega}_{j} t}), \tilde{c}^{in}_{j}=\hat{c}^{in}_{j}\exp({i\tilde{\omega}_{j}t})$, $\delta\tilde{d}_{j}=\delta\hat{d}_{j}\exp({i\omega_{m_j} t})$ and $\tilde{d}^{in}_{j}=\hat{d}^{in}_{j}\exp({i\omega_{m_j} t})$, the dynamics of the system reduce to the following linearized QLEs:
\begin{align}\label{eq:4}
\begin{split}
\frac{d}{dt}\delta\tilde{c}_{j}=&i(G_{j_-}+G_{j_+}e^{-2i\omega_{m_j} t})\delta\tilde{d}_{j}\\&+i(G_{j_+}+G_{j_-}e^{2i\omega_{m_j} t})\delta\tilde{d}^\dagger_{j} -\frac{\kappa_{j}}{2}\delta\tilde{c}_{j} \\&+\sqrt{\kappa_{j}}\tilde{c}^{in}_{j} +\Lambda e^{i\phi }e^{-i(\omega_{p}-\tilde{\omega}_{j}-\tilde{\omega}_{k})t}\delta\tilde{c}^\dagger_{k},\\
\frac{d}{dt}\delta\tilde{d}_{j}=&i(G_{j_-}+G_{j_+}e^{2i\omega_{m_j} t})\delta\tilde{c}_{j}  \\&+i(G_{j_+}+G_{j_-}e^{2i\omega_{m_j} t})\delta\tilde{c}^\dagger_{1}\
-\frac{\gamma_{j}}{2}\delta\tilde{d}_j\\&+\sqrt{\gamma_{j}}\tilde{d}^{in}_{j}.
\end{split}
\end{align} 
Here $G_{j_{\pm}}=g_{j} c^{s}_{j_{\pm}}$ are the effective coupling rates, where we assume $G_{j_{\pm}} > 0$. {In the weak coupling regime ($G_{j_{\pm}} < \kappa_j$) and under the resolved sideband (good cavity) limit ($\kappa_j \ll \omega_{m_j}$), the steady-state cavity field amplitudes $c^{s}_{j_\pm}$ can, without loss of generality, be taken as real. Under weak coupling, the detuning of the two-tone fields can be well approximated by $\Delta_{j_-} = +\omega_{m_j}$ for anti-Stokes scattering and $\Delta_{j_+} = -\omega_{m_j}$ for Stokes scattering~\cite{b53}. Moreover, the frequency matching condition $\omega_p = \tilde{\omega}_1 + \tilde{\omega}_2$ can be satisfied by appropriately tuning the NDOPA pump frequency. In the weak and resolved sideband parametric regime i.e., $\Lambda, G_{j_{\pm}}, \kappa_j \ll \omega_{m_j}$, and with $\gamma_j \ll \kappa_j \ll \omega_{m_j}$, the rotating wave approximation (RWA) allows us to neglect terms with counter-rotating factors $\exp(\pm 2i\omega_{m_j} t)$ in~\eqref{eq:4}. Based on the linearized quantum Langevin equations under RWA, squeezing the two-mode nanomechanical systems is achieved via cumulative quantum state transfer. This originates from the squeezing induced in the nanomechanical resonators through two-tone driving and the intracavity squeezing in the NDOPA, with beam-splitter-type interactions facilitated by the red-detuned laser drives.}
         
Furthermore, we define the pertinent quadrature operators to quantify quantum squeezing as $\delta {x}_{O_j}=(\delta\tilde{O}^\dagger_{j}+\delta\tilde{O}_{j})/\sqrt{2}$ and $\delta {y}_{O_j} = i(\delta\tilde{O}^\dagger_{j} -\delta\tilde{O}_{j})/\sqrt{2}$. Similarly, the corresponding input noise quadrature operators are given by ${x}^{in}_{O_j}=(\tilde{O}^{in,\dagger}_{j} +\tilde{O}^{in}_{j})/\sqrt{2}$ and ${y}^{in}_{O_j} = i(\tilde{O}^{in,\dagger}_{j}-\hat{O}^{in}_{j})/\sqrt{2}$. Thus, the dynamical equations in terms of quadrature fluctuations under the RWA of~\eqref{eq:4} can be written in a compact form {as an operator differential equation (ODE)}:
\begin{align}\label{eq:5}
\dot {\mathcal{U}}(t)=\mathcal{W}\mathcal{U}(t)+\mathcal{V}(t),
\end{align}
with $ \mathcal{U}(t)^T= [\delta{ {x}}_{c_1},\delta{ {y}}_{c_1},\delta{ {x}}_{c_2},\delta{{y}}_{c_2},\delta{ {x}}_{d_1},{ {y}}_{d_1},\delta{ {x}}_{d_2},\delta{ {y}}_{d_2}]$ is the vector of quadrature fluctuation operators, $\mathcal{V}(t)^T =[\sqrt{\kappa_{1}}{x}^{in}_{c_1}, \sqrt{\kappa_{1}}{y}^{in}_{c_1},\sqrt{\kappa_{2}}{x}^{in}_{c_2}, \sqrt{\kappa_{2}}{y}^{in}_{c_2},\sqrt{\gamma_{1}}{x}^{in}_{d_1}, \sqrt{\gamma_{1}}{y}^{in}_{d_1},\\ \sqrt{\gamma_{2}}{x}^{in}_{d_2}, \sqrt{\gamma_{2}}{y}^{in}_{d_2}]$ is the vector of fluctuating quadrature noise operators, and $\mathcal{W}$ is an $8\times 8$ drift matrix given by

\begin{align}\label{eq:6}
\mathcal{W}=
\begin{bmatrix}
-\frac{\kappa}{2} & 0 &C & S&0 & -A&0 & 0\\
0 & -\frac{\kappa}{2}& S &-C & B & 0 &0 & 0\\
C & S & -\frac{\kappa}{2} &0 & 0 & 0&0 & -A\\
S& -C &0 & -\frac{\kappa}{2}& 0&0&B & 0\\
0 & -A & 0 & 0&-\frac{\gamma} {2} &0&0 & 0\\
B & 0& 0 &0& 0&-\frac{\gamma} {2}&0 & 0\\
0& 0 &0 & -A&0& 0&-\frac{\gamma} {2}&0 \\
0 & 0 &B & 0& 0& 0 &0 & -\frac{\gamma} {2} 
\end{bmatrix}.  
\end{align}
Here, for simplicity, we set $G_{1\pm}=G_{2\pm}=G_{\pm}$, {(i.e., $P_{1\pm}=P_{2\pm}=P_{\pm}$)}, $\gamma_{1}=\gamma_{2}=\gamma$, $\kappa_{1}=\kappa_{2}=\kappa$. The uppercase letters are represented as $A=G_{-}-G_{+}$, $B=G_{-}+G_{+}$, $C=\Lambda\cos{\phi}$ and $S=\Lambda\sin{\phi}$.   

{The eigenvalues of the drift matrix $\mathcal{W}$ of~\eqref{eq:6} can be evaluated through the eigenvalue equation
\begin{equation}\label{eq:7}
\det{(\mathcal{W}-\lambda I)}=0    
\end{equation}
with $\lambda$ is eigenvalue and $I$ is unit matrix of $8\times 8$ dimension. When the system is stable and attains its steady state as $t\to \infty$, all the eigenvalues of the drift matrix $\mathcal{W}$  have negative real parts i.e., $\textrm{Re}(\lambda_l)<0$. Moreover, the stability conditions can be derived from the Ruith-Hurith stability criterion (RHSC)~\cite{b54}. The derivation of RHSC for the system from the characteristics polynomial equation is given in Appendix~\ref{appA}.}   

{Furthermore, the quadrature fluctuations of the system are described by a zero-mean multivariate Gaussian distribution. Accordingly, the equation of motion for the state of the system can be characterized through the covariance matrix (CM) making use of the ODE in \eqref{eq:5} as
\begin{equation}\label{eq:8}
\sigma(t)=\big<\mathcal{U}(t)\mathcal{U}^{T}(t)\big>,
\end{equation}
where the detailed derivation of the covariance matrix dynamics (time evolution of $\sigma(t)$) is provided in Appendix~\ref{appB}. For a stable system we evaluate the solution of the covariance matrix at the steady state employing the Lyapunov equation~\cite{b55} given by:}
\begin{equation}\label{eq:9}
\mathcal{W}\sigma + \sigma \mathcal{W}^{T} = -\mathcal{D}.
\end{equation}
This equation is linear in $\sigma$ and, in principle, can be solved analytically. However, the exact analytical expression is cumbersome and will not be presented here. {Moreover, the comparison of the stability conditions using RHSC for and the numerical integration of the time evolution of $\sigma(t)$ for the system at its steady state using numerical parameters that satisfy the condition are discussed in Appendix~\ref{appC}. In addition, the input noises are stationary and delta-correlated, it is appropriate to define the diffusion matrix as
\begin{equation}\label{eq:10}
\mathcal{D}\delta(t-t')= \big< \mathcal{V}(t)\mathcal{V}^T(t')\big>
\end{equation}
Here we consider the diffusion matrix $\mathcal{D}$ as a symmetric matrix like the covariance matrix $\sigma$. Therefore, the diffusion matrix $\mathcal{D}$, which is derived from \eqref{eq:10} by using the input noise delta-correlation functions and it takes the form}  
\begin{align}\label{eq:11}
\mathcal{D} = \text{diag} \big[ \kappa(n_c + \tfrac{1}{2}),\, \kappa(n_c + \tfrac{1}{2}),\, \kappa(n_c + \tfrac{1}{2}),\, \kappa(n_c + \tfrac{1}{2}), \notag \\
\gamma(n_m + \tfrac{1}{2}),\, \gamma(n_m + \tfrac{1}{2}),\, \gamma(n_m + \tfrac{1}{2}),\, \gamma(n_m + \tfrac{1}{2}) \big],
\end{align}
where we assume equal thermal bath temperatures, $T_1=T_2=T$ so that $n_{c_j(m_j)}=n_{c(m)}$.

\begin{figure*}
    \centering
    \includegraphics[width=\linewidth]{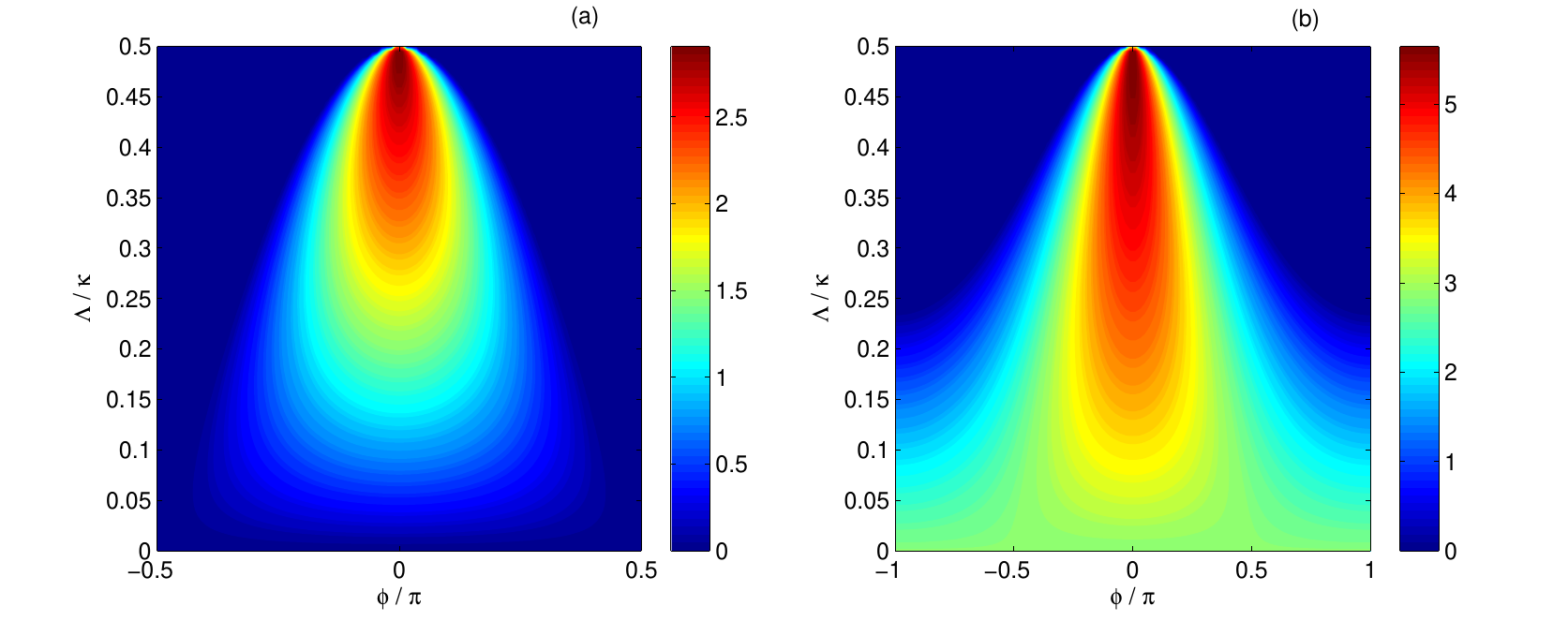}
    \caption{Density plots of the degree of two-mode quadrature squeezing of mechanical modes $S^{2}_m$(dB) as a function of normalized parametric coupling coefficient $(\Lambda/\kappa)$ versus phase $(\phi/\pi)$ of NDOPA for (a) $P_{_+} / P_{_-} =0 $, and (b) $P_{_+} / P_{_-} =0.1$ corresponding to quadrature squeezing $\big<(\delta x_{d})^2\big>$ when {$P_{_-}=10$ nW}, and {$T=10$ mK}. The other parameters are found in the main text.}\label{fig:2} 
\end{figure*}

\section {Quantifying squeezing}
\label{sec:Quantifying squeezing}

{In this section, we quantify two-mode squeezing without analyzing the detection mechanism, such as the reliable approach based on analyzing the eigenvalues of the covariance matrix~\cite{b56}. Two-mode squeezing can be characterized by several criteria, including the Duan bound criterion~\cite{b41} and methods that seek the maximally hybrid quadrature~\cite{b57,b58}, among others. Specifically, here to quantify the two-mode squeezing in bipartite systems, we employ noncommuting collective quadrature fluctuation operators used in several previous works~\cite{b44, b59,b60, b61}}. We define collective quadrature operators $\delta{x}_{c}$ and $\delta{y}_{c}$ for cavity modes while $\delta{x}_{d}$ and $\delta{y}_{d}$ for mechanical modes are given by 
$ \delta{x}_{c}=(\delta\tilde{c}^\dagger_{1}+\delta\tilde{c}_{1}+\delta\tilde{c}^\dagger_{2}+\delta\tilde{c}_{2})/\sqrt{2},	\delta{y}_{c}=i(\delta\tilde{c}^\dagger_{1}-\delta\tilde{c}_{1}+\delta\tilde{c}^\dagger_{2}-\delta\tilde{c}_{2})/\sqrt{2}, \delta{x}_{d}=(\delta\tilde{d}^\dagger_{1}+\delta\tilde{d}_{1}+\delta\tilde{d}^\dagger_{2}+\delta\tilde{d}_{2})/\sqrt{2}, \delta{y}_{d}=i(\delta\tilde{d}^\dagger_{1}-\delta\tilde{d}_{1}+\delta\tilde{d}^\dagger_{2}-\delta\tilde{d}_{2})/\sqrt{2}$ .

{The quantum average we are using does not indicate the ground, thermal equilibrium, and coherent states. It implies a Gaussian quantum state, in which the quantum fluctuations in an optomechanical cavity are typically evaluated in the system’s linearized steady state. This Gaussian quantum state includes the effects of environmental thermal noise, coherent laser drives (coherent laser amplitude and noise), and quantum back-action from radiation pressure.} Accordingly, the mean square fluctuations (variances or second moments) of the two-mode quadrature fluctuation operators in {the system's linearized steady state are derived} as {
\begin{align}\label{eq:12}  
\begin{split} 
\big<(\delta x_{c})^2\big>=&\big<(\delta{x}_{c_1}+\delta{x}_{c_2})^2\big>= \sigma_{11}+ \sigma_{33}+2 \sigma_{13},\\ 
\big<(\delta y_{c})^2\big>=&\big<(\delta{y}_{c_1}+\delta{y}_{c_2})^2\big>= \sigma_{22}+ \sigma_{44}+2\sigma_{24}, \\ 
\big<(\delta x_{d})^2\big>=&\big<(\delta{x}_{d_1}+\delta{x}_{d_2})^2\big>= \sigma_{55}+ \sigma_{77}+2 \sigma_{57},  \\
\big<(\delta y_{d})^2\big>=&\big<(\delta{y}_{d_1}+\delta{y}_{d_2})^2\big>=\sigma_{66}+ \sigma_{88}+2\sigma_{68}. 
\end{split}
\end{align}}

\begin{figure*}
    \centering
    \includegraphics[width=\linewidth]{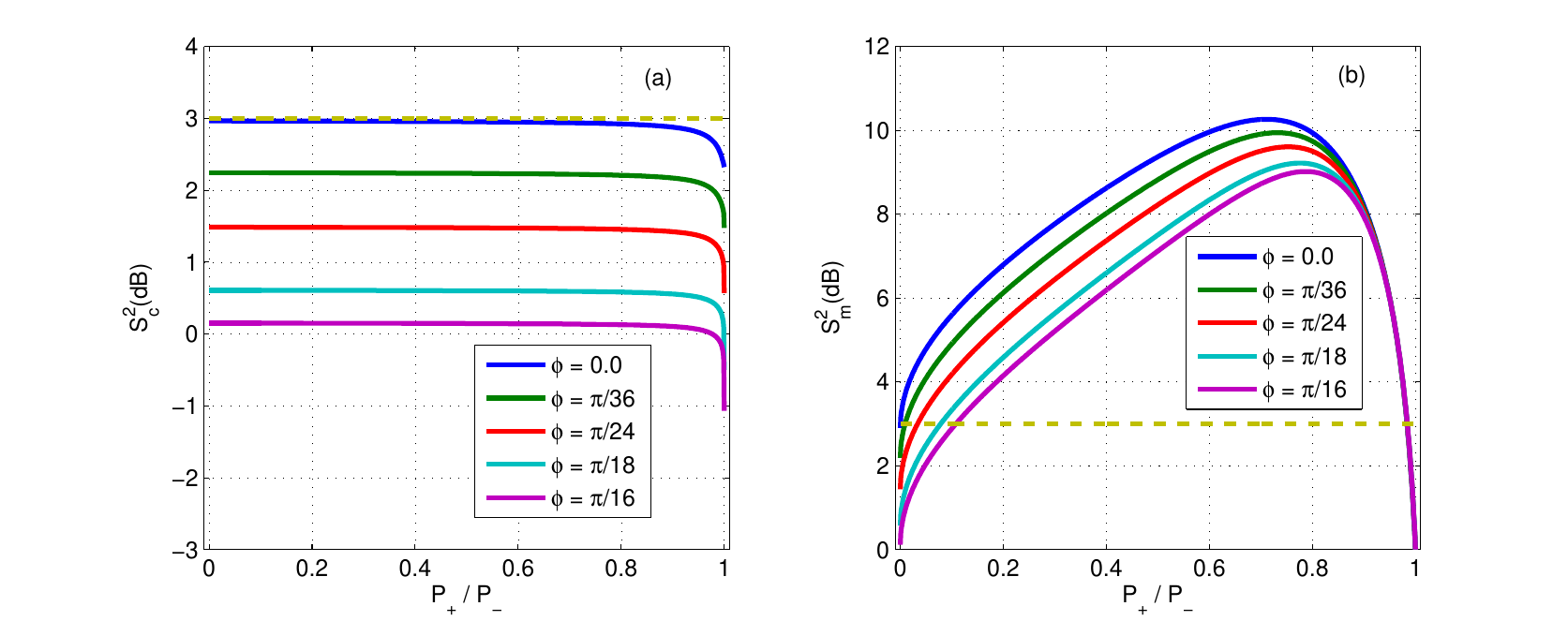}
    \caption{Plots of the degree of (a) two-mode optical squeezing $S^{2}_c$(dB) corresponding to quadrature squeezing $\big<(\delta y_{c})^2\big>$ and (b) two-mode mechanical squeezing $S^{2}_m$(dB) corresponding to quadrature squeezing $\big<(\delta x_{d})^2\big>$ versus the ratio of blue-to-red detuning powers $(P_{_+} / P_{_-})$ for the different parametric phases of NDOPA $\phi$ when the red-detuned laser drive {$P_{_-}=3$ nW}, and parametric coupling coefficient of NDOPA $\Lambda=0.49\kappa$. The other parameters are the same as Fig.~\ref{fig:2}}.
    \label{fig:3}
\end{figure*}

\begin{figure*}
    \centering
    \includegraphics[width=\linewidth]{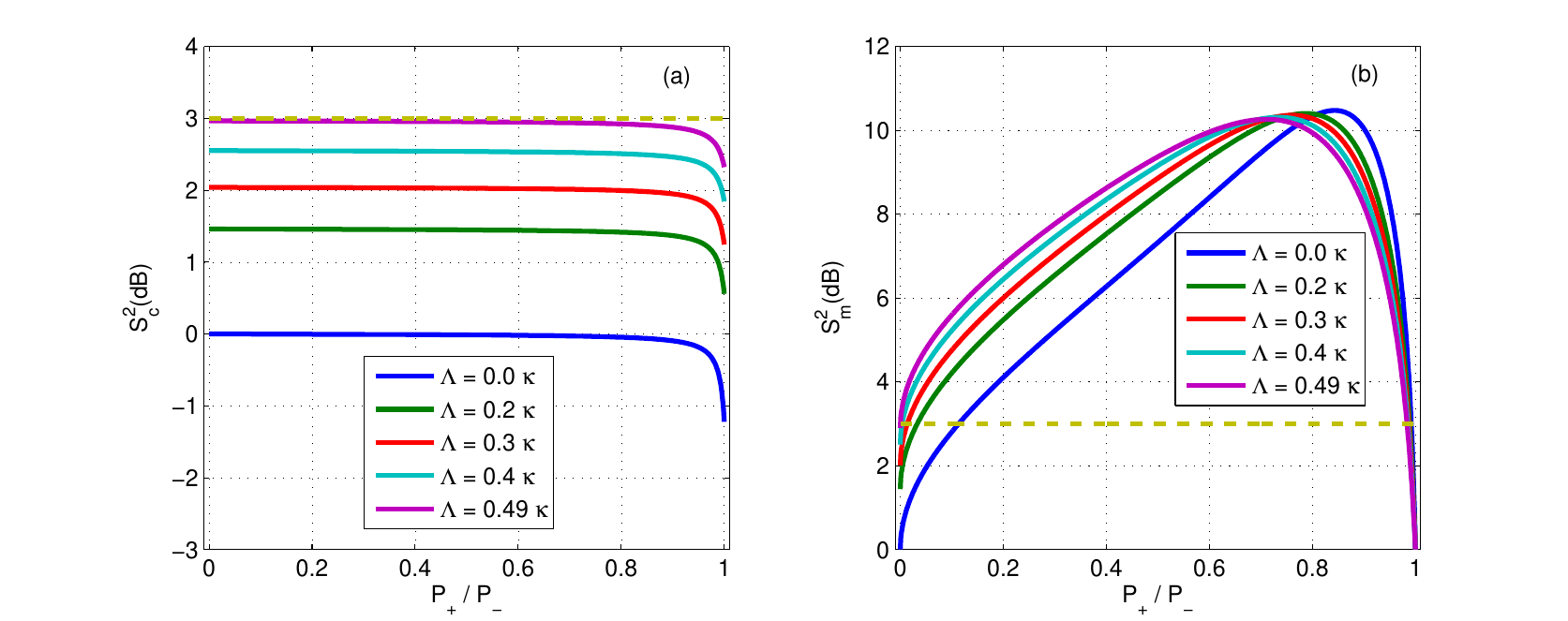}
    \caption{Plots of the degree of two-mode squeezing in (a) optical modes  $S^{2}_c$(dB) and (b) mechanical modes $S^{2}_m$(dB) versus the ratio of blue-to-red detuning powers $(P_{_+} / P_{_-})$ for different parametric coupling coefficient $\Lambda$ of NDOPA when parametric phase of NDOPA $\phi=0$ and the red-detuned laser drive {$P_{_-}=3$ nW}. The other parameters are the same as Fig.~\ref{fig:2}.}
    \label{fig:4}
\end{figure*}

Hence, the mechanical and cavity modes bipartite systems are in a two-mode squeezed state provided that one of the collective quadrature variances is below the shot-noise level. This phenomenon occurs when the following inequalities, $\big<(\delta x_{c})^2\big><1$ or $\big<(\delta y_{c})^2\big> <1$ for cavity modes, and $\big<(\delta x_{d})^2\big><1$ or $\big<(\delta y_{d})^2\big> <1$ for mechanical modes are satisfied. As a result, the Heisenberg uncertainty relations $\big<(\delta x_{c})^2\big> \big<(\delta y_{c})^2\big> \ge 1$ and $\big<(\delta x_{d})^2\big> \big<(\delta y_{d})^2\big> \ge 1$ hold. The degree of two-mode optical (mechanical) squeezing {$S^2_{c}(S^2_{m})$} can be expressed in units of decibel (dB)~\cite{b8} as { 
\begin{equation}\label{eq:13}
\begin{split}
S^2_{c(m)}(dB)=-10 \log_{10} \bigg[\frac{\big<(\delta R)^2\big>}{\big<(\delta R)^2 \big>_0}\bigg],
\end{split}
\end{equation}
where $\big<(\delta R)^2 \big>_0=1$ is the collective quadrature variance of the quantum shot-noise (thermal noise level of ground state) for cavity modes (mechanical modes) and $\delta R=\delta x_{c}$ or $\delta y_{c}$ for cavity modes and $\delta R=\delta x_{d}$ or $\delta y_{d}$ for mechanical modes. Thus, one of the collective quadrature variances of the cavity and mechanical modes could exhibit two-mode squeezing.}

\begin{figure*}
    \centering
    \includegraphics[width=\linewidth]{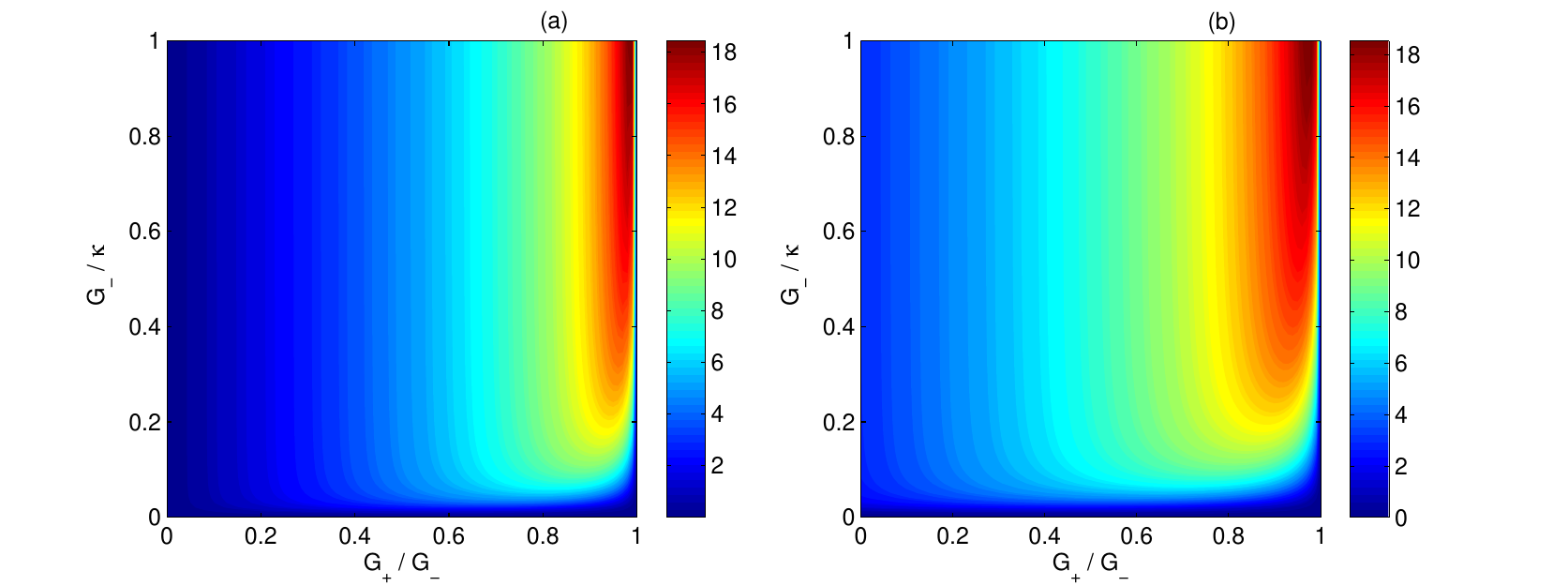}
    \caption{Density plots of the degree of two-mode mechanical squeezing $S^{2}_m$(dB) as a function of the the normalized red-tone driving strength $G_{_-}$ versus the blue-tone driving strength $G_{_+}$ (normalized by red-tone driving strength $ G_{_-}$) for (a) $\Lambda=0$, and (b) $\Lambda=0.49 \kappa$ when $\phi=0$. The other parameters are the same as Fig.~\ref{fig:2}.} \label{fig:5}
\end{figure*}

{In the present work, we investigate whether the quantum shot noise limit of 3 dB can be surpassed via two-mode squeezing between optical cavity modes and mechanical modes. In contrast, classical squeezing also known as thermomechanical noise squeezing can be used to suppress thermal noise. However, the reduction of thermal noise in one of the quadratures is fundamentally limited: without mechanical parametric amplification, the maximum achievable squeezing is limited to 3 dB below the thermal noise level, while with mechanical parametric amplification, it can be extended up to 6 dB~\cite{b62,b63,b64}. For the numerical simulations used to quantify the level of squeezing, we adopt typical experimentally feasible parameters of a microwave optomechanical system, as reported in~\cite{b38}, which aligns well with the operational regime of the current model under investigation. The mechanical resonator frequency is set to $\omega_{m_j}= 2\pi\times 3.6 \times 10^6$ Hz, while the frequency of the cavity field $\omega_{j}=2\pi \times 6.23 \times 10^{9}$ Hz, and the blue-detuned laser frequency as $\omega_{j_+}=\omega_{j}+\omega_{m_j}$. The cavity decay rate is $\kappa = 2\pi \times 450 \times 10^3$ Hz, which corresponds to $\kappa = 0.125\omega_{m_j}$, and the mechanical damping rate is $\gamma = 2\pi \times 3$ Hz, yielding $\gamma \approx 6.67 \times 10^{-6}\kappa$. The single-photon optomechanical coupling strength is taken as $g_j = 2\pi \times 36$ Hz and the mechanical thermal bath at $T = 10$ mK. Furthermore, laser drives applied at the red sideband frequency $\omega_{j_-} = \omega_{j} - \omega_{m_j}$ facilitate quantum ground state sideband cooling. This cooling technique enhances the system's feasibility of generating quantum mechanical squeezing.}

We evaluate the elements of CM from the numerical solution of the Lyapunov equation~\eqref{eq:9}. We quantify the degree of two-mode squeezing utilizing~\eqref{eq:13}, whereas the single-mode squeezing is directly determined from the CM diagonal elements corresponding to quadrature variances. Based on the numerical results, there is single mode displacement quadrature squeezing due to the reservoir engineering mechanism~\cite{b28}, that is, $\big<(\delta x_{d_1})^2\big>=\big<(\delta x_{d_2})^2\big><0.5$ but there is no quadrature squeezing corresponding to momentum, that is, $\big<(\delta y_{d_1})^2\big>=\big<(\delta y_{d_2})^2\big>>0.5$. Moreover, the individual quadrature squeezing of cavity modes due to NDOPA are all above the shot noise limit, i.e., $\big<(\delta x_{c_j})^2\big>>0.5$ and$\big<(\delta y_{c_j})^2\big> >0.5$~\cite{b25}. {On the other hand, the two-mode optical squeezing is due to the collective quadrature variance in phase sum, $\big<(\delta y_{c})^2\big><1$. In contrast, the two-mode mechanical squeezing emerges from the collective quadrature variance in position sum, $\big<(\delta x_{d})^2\big><1$. Thus, we focus on generating the two-mode squeezing due to these quadrature variances in dB units.} 

In Fig.~\ref{fig:2}, we plot the two-mode squeezing as a function of the parametric coupling coefficient and phase. The effects of the red-detuning laser drive are clearly shown in Fig.~\ref{fig:2}~(a) for {$P_-=10$ nW} without blue-detuning tones. Thus, the two-mode mechanical squeezing is generated due to the quantum state transfer of correlated photons of NDOPA~\cite{b25}. Accordingly, the optimal squeezing {2.95} dB below the 3 dB level is attained at $\phi=0$ and {$\Lambda=0.495\kappa$}. Moreover, at higher values of the parametric coupling coefficient, squeezing is generated within a small range of the parametric phase compared to the lower limit of the parametric coupling coefficient from $-\pi/2$ to $\pi/2$. On the other hand, the joint effect of parametric amplification with red and blue detuning for $P_{_+} / P_{_-} = 0.1$ for {$P_{_-} =10$ nW} is shown in Fig.~\ref{fig:2}~(b). Here, with two pairs of two-tone laser driving effects, we attain an optimal 5.7 dB at $\phi=0$ and $\Lambda=0.485\kappa$. In addition, at higher values of the parametric coupling coefficient, squeezing is generated for small ranges of the parametric phase, while at the lower parametric coupling coefficient, it can extend from $-\pi$ to $\pi$. When we compare the plots Fig.~\ref{fig:2}~(a) and Fig.~\ref{fig:2}~(b), it is possible to beat the 3 dB squeezing limit and modulate the squeezing for a wide range of parametric phases with two-tone lasers.

\begin{figure*}
    \centering
    \includegraphics[width=\linewidth]{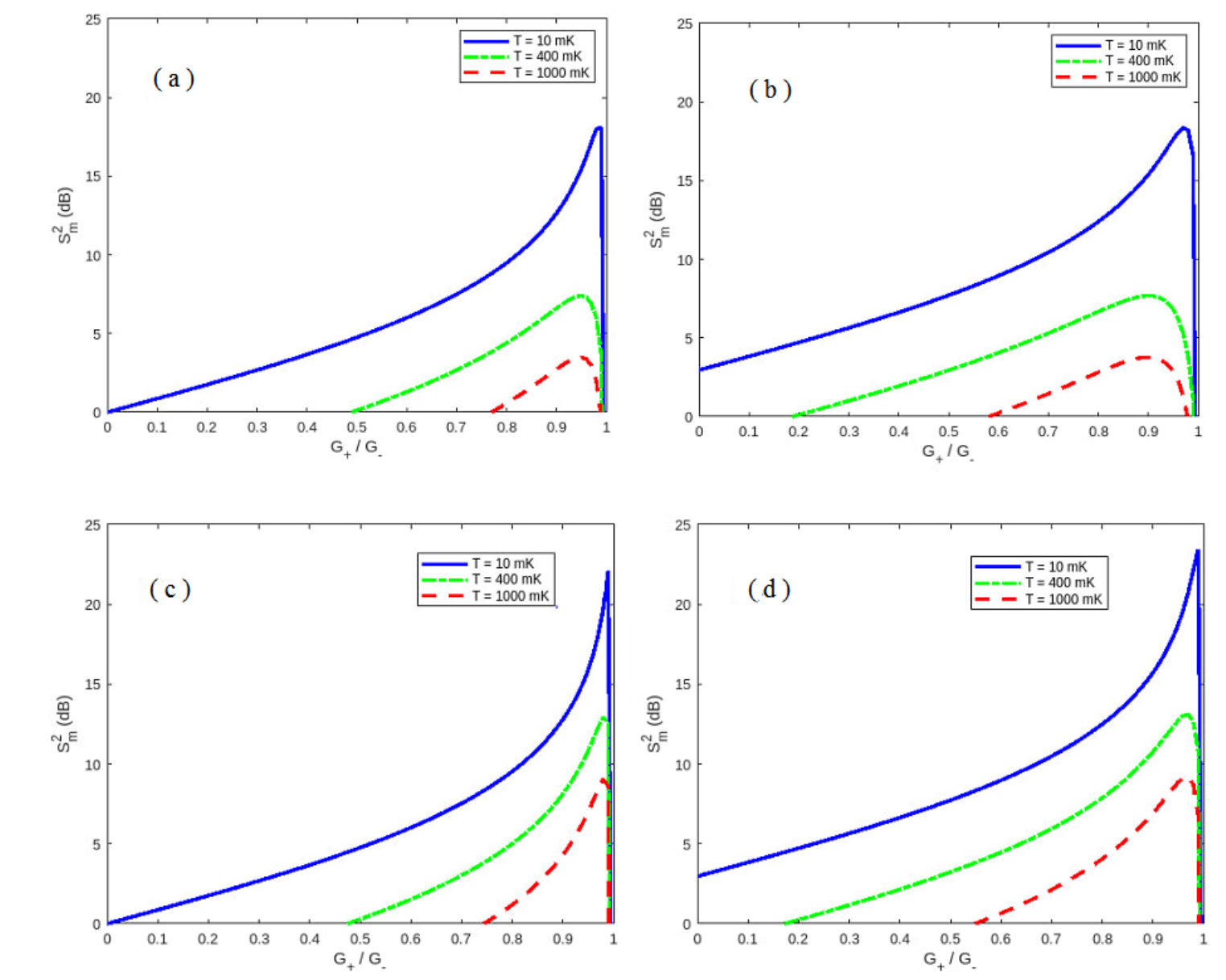}
    \caption{Plots of the degree of two-mode mechanical squeezing $S^{2}_m$(dB) versus the ratio of blue-to-red detuning lasers coupling strength $(G_{_+} / G_{_-})$ for different temperature $T$ (mK) of mechanical baths when $\phi=0$, $G_{_-}/\kappa=1$ for (a) $\Lambda =0$, $\gamma =6.67 \times 10^{-6} \kappa$ (b) $\Lambda =0.49\kappa$, $\gamma =6.67 \times 10^{-6} \kappa$ (c) $\Lambda =0$, $\gamma =0.667 \times 10^{-6} \kappa$ and (d) $\Lambda =0.49\kappa$, $\gamma =0.667 \times 10^{-6} \kappa$. The other parameters are the same as Fig.~\ref{fig:2}.}
    \label{fig:6}
\end{figure*}

Furthermore, we have investigated the impacts of two pairs of two-tone laser driving on two-mode squeezing in cavity modes and mechanical mode bipartite subsystems, as shown in Fig.~\ref{fig:3}, for different parametric phases $\phi$ at red-detuning laser power {$P_{_-}=3$ nW}. In Fig.~\ref{fig:3}~(a), the squeezing in cavity modes cannot beat the conventional 3 dB limit. Additionally, the degree of optical squeezing decreases as the phases increase from $\phi=0$ to $\phi=\pi/16$. It is nearly a constant for all ranges of $P_{_+} / P_{_-}$ except near its optimal value $P_{_+} / P_{_-}=1$ where the optical modes squeezing becomes reduced. However, the two pairs of two-tone laser drives significantly affect the two-mode mechanical squeezing, as depicted in Fig.~\ref{fig:3}~(b). In this plot, as the parametric phase increases, the squeezing decreases. For a lower ratio $P_{_+} / P_{_-}$ near 0.1, the mechanical squeezing is above the 3 dB for parametric phases $\phi=0, \pi/36, \pi/24$. In contrast, the mechanical squeezing is lower than the level 3 dB for $\phi=\pi/18, \pi/16$. Moreover, by increasing the ratio $P_{_+} / P_{_-}$, we achieve mechanical squeezing above 3 dB while the squeezing grows to an optimal value corresponding to each parametric phase. In addition, near the optimal value of the ratio $P_{_+} / P_{_-}$, the effect of the parametric effect is negligible, and the squeezing corresponding to each parametric phase is nearly identical, converging to the vacuum noise level at $P_{_+} / P_{_-}=1$. The configuration leads to balanced driving conditions in back-action evading~\cite{b28}. For this limit, all back-action noise is flowed to the momentum $(\delta {y}_{d_j})$ quadrature, leaving the displacement $(\delta {x}_{d_j})$ quadrature unaffected, so that neither of the quadratures is cooled~\cite{b41}.

\begin{figure*}
    \centering
    \includegraphics[width=\linewidth]{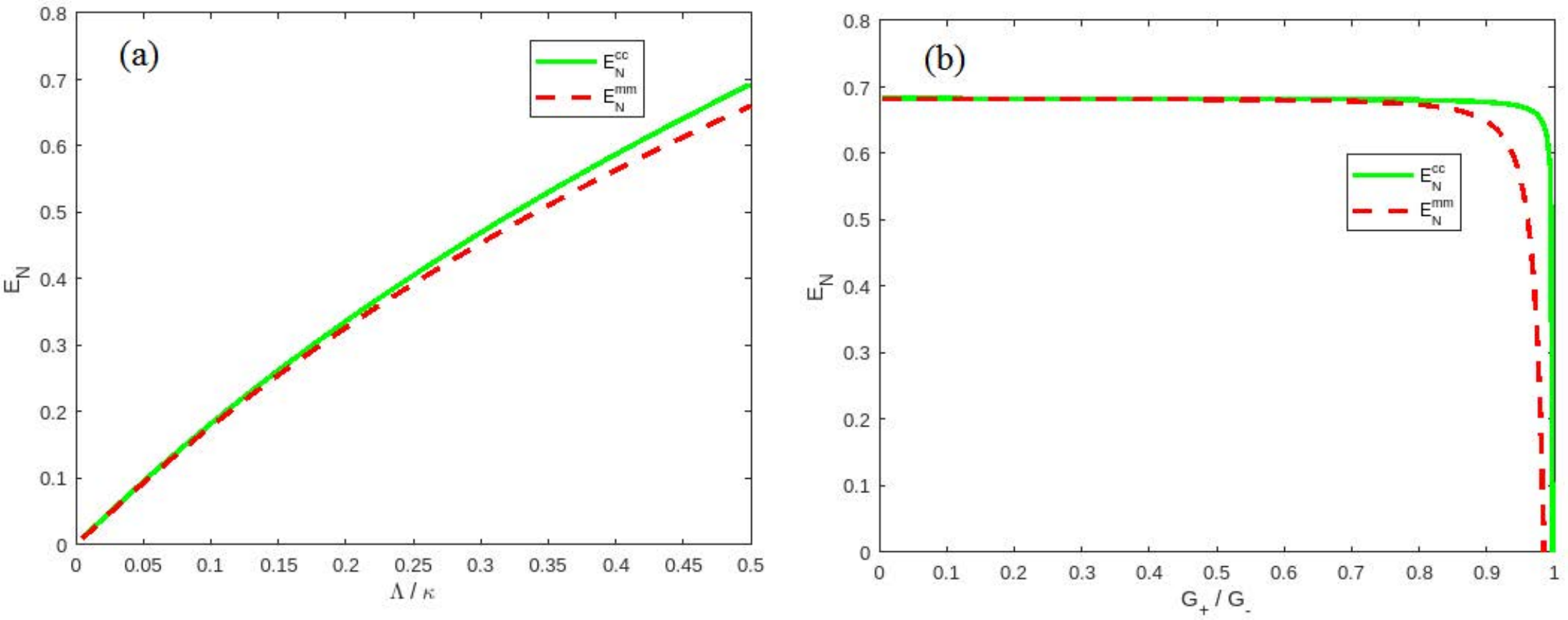}
    \caption{Plots of the degree of entanglement in optical modes  $E^{cc}_N$ and mechanical modes $S^{mm}_N$ versus (a) normalized parametric coupling coefficient ($\Lambda/\kappa$) for $G_{_-}=0.1\kappa$, $G_{_+} / G_{_-}=0.1$, $\phi=0$ and (b) the ratio of blue-to-red detuning coupling strengths $(G_{_+} / G_{_-})$ for $G_{_-}=1\kappa$ , $\Lambda=0.49\kappa$, $\phi=0$. The other parameters are the same as Fig.~\ref{fig:2}.}
    \label{fig:7}
\end{figure*}

The plots in Fig.~\ref{fig:4} display the results of the two-mode squeezing in cavity and mechanical modes for different parametric coupling coefficients. Thus, in Fig.~\ref{fig:4}~(a), the squeezing in cavity modes is below the 3 dB limit even though the non-linear gain increases to its threshold value. Furthermore, the degree of optical squeezing increases as the parametric coupling coefficient increases from $\Lambda=0$ to $\Lambda=0.49\kappa$. For $\Lambda \ne 0$ the optical modes squeezing has a constant value for all ranges of $P_{_+} / P_{_-}$ except near its optimal ratio $P_{_+} / P_{_-}=1$ where it shows reduction to some extent. On the other hand, in the plot of Fig.~\ref{fig:4}~(b), the two pairs of two-tone laser drives exhibit considerable effects on the degree of two-mode mechanical squeezing. Accordingly, the mechanical squeezing becomes stronger as the parametric coupling coefficient increases to a certain limit of $P_{_+} / P_{_-}$ approaching 1. For example, at the lower ratio of $P_{_+} / P_{_-}$ near 0.1, the two-mode mechanical squeezing is below the 3 dB level for parametric gain of $\Lambda=0$. However, this squeezing breaks the 3 dB limit for other nonzero gains. Moreover, by increasing the ratio $P_{_+} / P_{_-}$, the mechanical squeezing becomes very strong until an optimal value corresponding to each parametric coupling coefficient decreases. Moreover, the optimal squeezing values corresponding to higher parametric coupling coefficients are lower than those of the lower parametric coupling coefficient of NDOPA. This occurs for a relatively smaller ratio $P_{_+} / P_{_-}$ value. This phenomenon is caused by a competitive quantum state transfer effect of single-mode squeezing being more dominant than the two-mode squeezing of NDOPA. For $P_{_+} / P_{_-}$ near 1, the effect of the non-linear gain becomes less significant, and the squeezing corresponding to each parametric gain is nearly a similar value and attains the vacuum noise level.

In Fig.~\ref{fig:5}, the plots of two-mode mechanical squeezing as a function of the normalized red-detuned optomechanical coupling strength $G_{_-}/\kappa$ versus the ratio $G_{_+} / G_{_-}$ when $G_{_-} =\kappa$, $\phi=0$, for different parametric coupling coefficients in the weak coupling limit. In the plot Fig.~\ref{fig:5}~(a), we investigate the effects of two-tone laser driving with the corresponding red-detuning laser coupling strength. Thus, the displacement squeezing of each mechanical mode is completely transferred to generate two-mode nanomechanical squeezing. Consequently, we investigated that at $G_{_-}=0$ all ranges of $G_{_+} / G_{_-}$, near $G_{_+} / G_{_-}=0$ all ranges of $G_{_-}/\kappa$ in the parametric limit, and at $G_{_+} / G_{_-}=1$ the degree of squeezing overlaps the vacuum noise level. Moreover, we see that as the ratio $G_{_+} / G_{_-}$ increases, the degree of squeezing starts to beat the conventional limit of 3 dB for a very low coupling strength $G_{_-}/\kappa$ and a relatively far distance from the optimal ratio of $G_{_+} / G_{_-}$. The effect of the coupling strength $G_{_-}/\kappa$ is more visible for a higher ratio of $G_{_+} / G_{_-}$. In addition to this, for a given value of $G_{_-}/\kappa$, increasing the ratio of $G_{_+} / G_{_-}$ leads to a degree of squeezing toward the optimal value corresponding to the red-detuning coupling strength. However, for various $G_{_-}$, the optimal squeezing points have different points in the ratio $G_{_+} / G_{_-}$. This shift in optimal squeezing is a result of the reduction in purity of mechanical squeezing with an increase in the ratio $G_{_+} / G_{_-}$ and the heating process of the blue-detuned laser driving field~\cite{b28, b34}. The optimal squeezing happened at {$G_{_+} / G_{_-}=0.985$}  and $G_{_-}/\kappa=1$ with a level of {18.30 dB}. Furthermore, the joint effects of NDOPA and two-tone laser driving have been depicted in Fig.~\ref{fig:5}~(b). Here we see that when $G_{_-}/\kappa=0$ and near $0$ for all ranges of $G_{_+} / G_{_-}$ and for $G_{_+} / G_{_-}=1$ and near $1$ the degree of squeezing is similar as in the former scenario of Fig.~\ref{fig:5}~(a). However, for $G_{_-}/\kappa>0$ and $G_{_+} / G_{_-}=0$, the degree of squeezing becomes nonzero due to the pumped NDOPA. The remaining features of the plot in Fig.~\ref{fig:5}~(b) with similar features as the Fig.~\ref{fig:5}~(a) but with a slightly higher optimal squeezing level of {18.40 dB}, which occurs at {$G_{_+} / G_{_-}=0.975$} and $G_{_-}/\kappa=1$.

{In the plots of Fig.~\ref{fig:6}, we plot relevant characteristics of environmental thermal effects, parametric amplification, and the importance of mechanical quality factors due to mechanical damping rates. In Fig.~\ref{fig:6}~(a) and Fig.~\ref{fig:6}~(b), and Fig.~\ref{fig:6}~(c) and Fig.~\ref{fig:6}~(d), the mechanical damping rates are $\gamma=6.67 \times 10^{-6} \kappa $ and $\gamma=0.667 \times 10^{-6} \kappa $ respectively, corresponding to mechanical quality factors of $1.2\times 10^6$ and $12\times 10^6$. In the plots, it is clearly shown that bath temperature has a negative effect on squeezing. Specifically, for $\Lambda=0.49\kappa$ at $T=10$ mK and a lower $G_{_+} / G_{_-}$ ratio of 0.1, it is possible to beat the 3 dB limit. For this case, in the absence of NDOPA, squeezing is below the 3 dB limit. However, at other higher temperatures (400 mK and 1000 mK) far from the optimal ratio $G_{_+} / G_{_-}$, the corresponding squeezings are below the shot noise level. Moreover, improvement in the mechanical quality factor becomes an agent to increase the optimal degree of squeezing, for example in the absence of NDOPA from 18.11 dB in Fig.~\ref{fig:6}~(a) to 22.14 dB in Fig.~\ref{fig:6}~(c) for identical values of $G_{_+} / G_{_-}=0.99$. Thus, the effect of a better mechanical quality factor is to make the degree of mechanical squeezing more robust against temperature. Utilizing the parametric coupling coefficient near its threshold value $\Lambda=0.49\kappa$, in Fig.~\ref{fig:6}~(b) and Fig.~\ref{fig:6}~(d), the degree of squeezing at $G_{_+} / G_{_-}=0$, is nonzero as compared to Fig.~\ref{fig:6}~(a) and Fig.~\ref{fig:6}~(c). The presence of NDOPA slightly shifts up the optimal squeezing, and the degree of squeezing exhibits higher robustness against thermal fluctuations. Increasing the mechanical quality factor with NDOPA also improves the optimal degree of squeezing from 18.37 dB in Fig.~\ref{fig:6}~(b) at $G_{_+}/G_{_-}=0.97$ to 23.47 dB in Fig.~\ref{fig:6}~(d) at $G_{_+}/G_{_-}=0.99$. Furthermore, the robustness of the squeezing against temperature can be achieved within the system by increasing the ratio $G_{_+}/G_{_-}$ towards its optimal.}

{Furthermore, it is relevant to quantify the entanglement transfer from NDOPA to cavity modes and then induced to mechanical modes. Thus, in Fig.~(\ref{fig:7}), we plot the entanglement between bipartite systems (cavity-cavity entanglement $E^{cc}_N$ and mechanical-mechanical entanglement $E^{mm}_N$) using logarithmic negativity~\cite{b73}. In these plots the optimal cavity modes entanglement transfer to the mechanical modes does not break the coherent bound level~\cite{b41}, $E_N=\ln{2}\approx0.693$, which is equivalent to the two-mode squeezing 3 dB limit, as shown in Fig.~\ref{fig:3}~(a) and Fig.~\ref{fig:4}~(a). Specifically, in Fig.~\ref{fig:7}~(a) for $G_{_-}=0.1k$, the entanglement between the bipartite systems becomes more robust as the parametric coupling coefficient goes to its threshold limit. Moreover, the mechanical entanglement at higher parametric coupling coefficients becomes lower due to converting pure squeezed cavity modes to thermal squeezed states. The overcoming of such an effect is compensated by enhancing the red-detuning coupling strength. On the other hand, in Fig.~\ref{fig:7}~(b) near the parametric coupling coefficient threshold value of $\Lambda=0.49\kappa$, the entanglement between cavity modes and mechanical modes is constant or independent for a wide range of $G_{_+} / G_{_-}$. However, a higher ratio of $G_{_+} / G_{_-}$ limits the mechanical entanglement to be lower than the cavity modes' entanglement. In addition, as the ratio $G_{_+} / G_{_-}$ increased towards its stability limit, the entanglement in both bipartite systems drastically decreased to zero. This is due to the increasing heating effect of the blue detuning coupling strengths.}        

{While we quantify two-mode squeezing using the variances of collective quadrature operators~\eqref{eq:11}, we suggest that an alternative and often more robust method involves evaluating the symplectic eigenvalues of the covariance matrix. This approach, as emphasized in~\cite{b56}, offers a phase-independent measure of squeezing and can serve as an optimal figure of merit. In our scheme, the observed dependence of squeezing on the parametric phase $\phi$, particularly illustrated in Fig.~\ref{fig:2} and Fig.~\ref{fig:3}, is the result of the phase-sensitive structure of the quadrature operators used in homodyne-based detection. This behavior is analogous to the dependence of single-mode squeezing on the local oscillator phase in standard optical homodyne detection. Nonetheless, our approach remains experimentally meaningful since the covariance matrix fully characterizes the Gaussian steady state. All quadrature variances quantifying squeezing can be reconstructed via homodyne detection with appropriately chosen local oscillator phases. Future work may explore the symplectic eigenvalue criterion as a complementary measure to strengthen the detection strategy further.}

\section{Experimental feasibility of two-Mode squeezing implementation} \label{sec:Experimental feasibility}

\begin{figure}
    \centering
    \includegraphics[width=\linewidth]{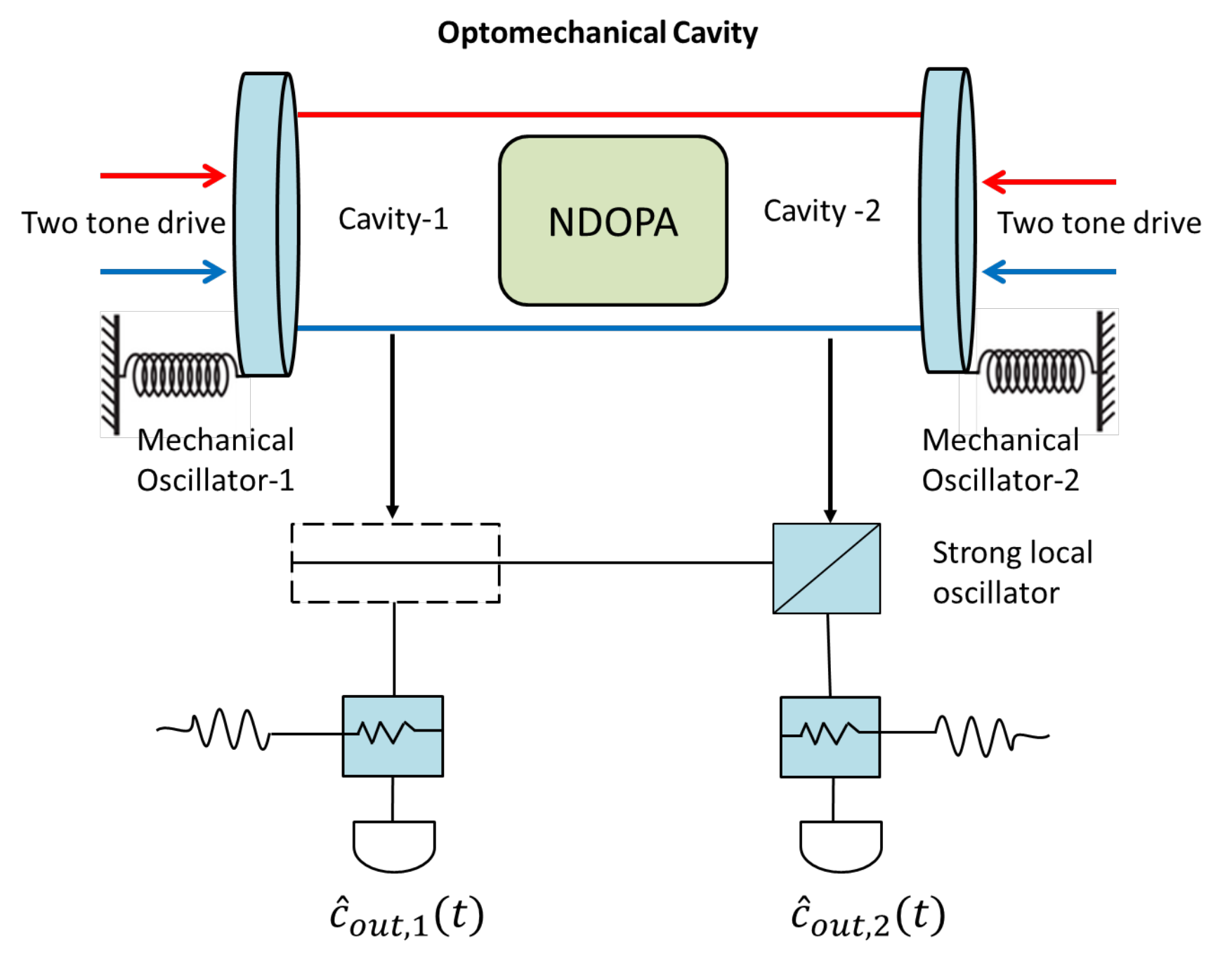}
    \caption{Schematic of experimental set up in an optomechanical cavity to verify two-mode mechanical squeezing using a standard homodyne scheme.}
    \label{fig:8}
\end{figure}

{The experimental configuration proposed in this work, comprising two mechanical oscillators coupled to two optical cavity modes within a single optomechanical cavity containing a nonlinear optical device (NDOPA), as illustrated in Fig.~\ref{fig:8}, is physically realizable with current technology. Each cavity mode is driven by two tones (red- and blue-detuned) to activate both beam-splitter and squeezing-type interactions. The output optical fields are monitored via standard homodyne detection to infer the two-mode mechanical squeezing.}

{Recent experiments have demonstrated key components of our setup. Multimode mechanical systems coupled to optical or microwave cavities have been realized, including two-membrane or two-end-mirror configurations~\cite{b65,b66}. Optical cavities supporting two distinct modes, either spatial, frequency, or polarization separated, are well-established in cavity QED and optomechanics platforms. Furthermore, embedding an NDOPA inside a cavity is a standard technique in continuous-variable quantum optics and has been used to generate strong two-mode optical squeezing~\cite{b67,b68}, although there are intracavity losses~\cite{b71}.}

{Two-tone driving has been widely employed to achieve squeezing in mechanical resonators, which is in ground state sideband cooling with a cryostat precooling technique and amplification processes~\cite{b38, b69, b70}. Similarly, balanced homodyne detection remains the most robust and accessible method for measuring quadrature fluctuations in optical fields, enabling the reconstruction of the covariance matrix~\cite{b72} and verification of squeezing.}

{While technical challenges remain, maintaining low phonon occupancy, strong optomechanical coupling, and high detection efficiency, state-of-the-art laboratories have individually demonstrated all of the required components. Therefore, our proposed protocol is feasible with currently available optomechanical and nonlinear optical systems.}

\section{ Conclusions} 
\label{sec:Conclusions}
We proposed and analyzed a scheme for enhancing two-mode squeezing beyond the conventional 3~dB limit in two nanomechanical oscillators coupled to a doubly resonant cavity. Our findings within the chosen parametric regime demonstrate that the degree of mechanical two-mode squeezing is highly sensitive to several key factors, including the pumping phase and parametric coupling coefficient of the NDOPA, the optomechanical coupling strengths of red- and blue-detuned lasers, as well as the temperature and damping rates of the mechanical baths. {The squeezing from the proposed scheme can be realized via homodyne detection for experimental parameters of the current-state-of art}. Moreover, the results emphasize the tunability of mechanical squeezing via various system parameters, making this scheme a promising platform for quantum technologies. Potential applications include quantum sensing and quantum information processing. Future work could extend this study by integrating the proposed mechanism into more complex quantum networks, particularly under varying environmental conditions.

\appendix

\section*{Appendix: SUPPLEMENTARY MATERIAL FOR STABILITY VERIFICATION} 

\section{Applicability of Routh-Hurwitz Stability Criterion (RHSC)}\label{appA}
\renewcommand{\theequation}{A.\arabic{equation}}\setcounter{equation}{0}
{This appendix demonstrates the applicability of the Routh–Hurwitz stability criterion (RHSC). In the main text,~\eqref{eq:5} represents the linearized Heisenberg–Langevin equation in terms of a continuous-variable Gaussian state with Hermitian quadrature fluctuation operators. Specifically, the RHSC ensures that all eigenvalues of the drift matrix have strictly negative real parts when the system reaches its stable state. From~\eqref{eq:7}, the RHSC corresponding to the characteristic equation can be obtained. Accordingly, the characteristic polynomial equation can be written in the form
\begin{equation}\label{eq:A.1}
(\lambda^4+s_1\lambda^3+s_2\lambda^2+s_3\lambda+s_4)^2=0,  
\end{equation}
where the coefficients $s_r$ ($r = 1,2,3,4)$ are given by
\begin{align*}
    &s_1=\gamma+\kappa, \\&s_2=\gamma\kappa+\frac{1}{4}(\gamma^2+\kappa^2-4\Lambda^2) +2(G^2_{-}-G^2_{+}),\\& s_3=(\gamma+\kappa)\big(\frac{1}{4}\gamma\kappa+G^2_{-}-G^2_{+}\big)- \gamma\Lambda^2, \\&s_4=(G^2_{-}-G^2_{+})\big(G^2_{-}-G^2_{+}+\frac{1}{2}\gamma\kappa\big)+\frac{1}{16}\gamma^2(\kappa^2-4\Lambda^2).
\end{align*}
The Routh–Hurwitz stability criteria impose the following stability conditions:
\begin{align}\label{eq:A.2}
\begin{split}
    s_r &> 0 \ \ \text{for} \ \ r = 1,2,3,4, \\
s_1 s_2 - s_3 &> 0,  \\
s_1 s_2 s_3 - s_3^2 - s_1^2 s_4 &> 0. 
\end{split}
\end{align}
These translate to the following parameter-dependent inequalities:}
\begin{align}\label{eq:A.3}
\begin{split}
h_{1}=& (G^2_{-}-G^2_{+})\big[G^2_{-}-G^2_{+}+\frac{1}{2}\gamma\kappa\big]\\&+\frac{\gamma^2}{16}(\kappa^2-4\Lambda^2) >0,\\
h_{2}=& \big[\gamma\kappa+\frac{1}{4}(\gamma^2+\kappa^2-4\Lambda^2) +2(G^2_{-}-G^2_{+})\big]\\&\times\big[(\gamma+\kappa)\big(\frac{\gamma\kappa}{4}+G^2_{-}-G^2_{+}\big)- \gamma\Lambda^2\big]\\&
-(\gamma+\kappa)h_{1}>0,	\\ 
h_{3}=&-\big[(\gamma+\kappa)\big(\frac{\gamma\kappa}{4}+G^2_{-}-G^2_{+}\big)- \gamma\Lambda^2\big]^2\\&+
(\gamma+\kappa)h_{2} >0. 
\end{split}
\end{align} 
We note that the stability conditions of the system do not depend on the parametric phase $\phi$. Moreover, considering the weak coupling limit where $\gamma \ll \kappa$, a sufficient condition for the stability of the system described by~\eqref{eq:A.3} is fulfilled when both $\Lambda < 0.5\kappa$ and $G_{+}<G_{-}$.

\section{Covariance Matrix Dynamics}\label{appB}
\renewcommand{\theequation}{B.\arabic{equation}}\setcounter{equation}{0}
{In this appendix, we derive the covariance matrix dynamics which can lead to the well-known Lyapunov equation at steady state. Equation~\eqref{eq:5} is an operator differential equation (ODE) in which the state vector $\mathcal{U}$ is not in $\mathbf{R}^n$ nor in $\mathbf{C}^n$. In addition, the creation or annihilation operators can be represented as infinite dimensional matrices. Hence, the Routh-Hurwitz stability criterion cannot be properly used in infinite dimensional systems; we employ the covariance matrix dynamics to overcome the challenge. From the ODE system of~\eqref{eq:5}, we can write:
\begin{align}\label{eq:B.1}
\begin{split}
    \dot {\mathcal{U}}(t)=&\mathcal{W}\mathcal{U}(t)+\mathcal{V}(t),\\
    \dot {\mathcal{U}}^{T}(t)=&\mathcal{U}^{T}(t)\mathcal{W}^{T}+\mathcal{V}^{T}(t).
\end{split}  
\end{align}
where we have used the identity $(XY)^T = Y^T X^T$. Further, utilizing the ODE system equation, we write the covariance matrix as $\sigma(t)=\big<\mathcal{U}(t)\mathcal{U}^{T}(t)\big> \in \mathbf{R}^{n\times n}$. Substituting \eqref{eq:B.1} into the time evolution of the covariance matrix $\sigma(t)$, 
\begin{equation}\label{eq:B.2}
    \dot \sigma (t)=\big< \mathcal{\dot U}(t)\mathcal{{U}}^{T}(t)\big> +\big< \mathcal{U}(t)\mathcal{\dot U}^{T}(t)\big>,
\end{equation}
we obtain 
\begin{align}\label{eq:B.3}
    \dot{\sigma}(t) &= \mathcal{W} \big\langle \mathcal{U}(t)\mathcal{U}^{T}(t) \big\rangle + \big\langle \mathcal{U}(t)\mathcal{U}^{T}(t) \big\rangle \mathcal{W}^T \nonumber \\
    &+ \big\langle \mathcal{V}(t)\mathcal{U}^{T}(t) \big\rangle + \big\langle \mathcal{U}(t)\mathcal{V}^{T}(t) \big\rangle.
\end{align}
This also implies
\begin{align}\label{eq:B.4}
    \dot{\sigma}(t)  &= \mathcal{W}\sigma(t) + \sigma(t)\mathcal{W}^T \nonumber \\ 
    &+ \big\langle \mathcal{V}(t)\mathcal{U}^{T}(t) \big\rangle + \big\langle \mathcal{U}(t)\mathcal{V}^{T}(t) \big\rangle.
\end{align}
Now, from the formal solution of~\eqref{eq:5}, we have
\begin{align}\label{eq:B.5}
\begin{split}
\mathcal{U}(t)=&e^{\mathcal{W}t}\mathcal{U}(0)+\int^t_0 e^{\mathcal{W}(t-t')}\mathcal{V}(t')dt', \\ 
\mathcal{U}^T(t)=&\mathcal{U}^T(0)e^{\mathcal{W}^Tt}+\int^t_0 \mathcal{V}^T(t') e^{\mathcal{W}^T(t-t')}dt',
\end{split}
\end{align}
where we used the identity $[\exp(X)]^T = \exp(X^T)$. Then using \eqref{eq:B.5} in the expression of the second line of \eqref{eq:B.4}, we obtain
\begin{align}\label{eq:B.6}
    \big< \mathcal{V}(t)\mathcal{{U}}^{T}(t)\big>&+\big< \mathcal{U}(t)\mathcal{V}^{T}(t)\big> =
   \big< \mathcal{V}(t)\mathcal{U}^T(0)\big>e^{\mathcal{W}^Tt}\nonumber \\ 
   &+\int^t_0 \big< \mathcal{V}(t)\mathcal{V}^T(t')\big> e^{\mathcal{W}^T(t-t')}dt' \nonumber \\ 
   &+e^{\mathcal{W}t}\big<\mathcal{U}(0)\mathcal{V}^{T}(t)\big>\nonumber \\ 
   &+\int^t_0 e^{\mathcal{W}(t-t')}\big<\mathcal{V}(t')\mathcal{V}^{T}(t)\big>dt'.
\end{align}
Further, using the fact that $\big< \mathcal{V}(t)\mathcal{U}^T(0)\big>=\big< \mathcal{U}(0)\mathcal{V}^T(t)\big>=0$ in \eqref{eq:B.6}, we have 
\begin{align}\label{eq:B.7}
    \big< \mathcal{V}(t)\mathcal{{U}}^{T}(t)\big> &+ \big< \mathcal{U}(t)\mathcal{V}^{T}(t)\big>\nonumber\\=&\int^t_0 \big< \mathcal{V}(t)\mathcal{V}^T(t')\big> e^{\mathcal{W}^T(t-t')}dt' \nonumber\\&+\int^t_0 e^{\mathcal{W}(t-t')}\big<\mathcal{V}(t')\mathcal{V}^{T}(t)\big>dt'\nonumber\\=&\mathcal{D}
    \end{align}
where $\mathcal{D} \in R^{n\times n}$ is the diffusion matrix. Substituting~\eqref{eq:B.7} into~\eqref{eq:B.4}, we obtain the time evolution of the covariance matrix as:
\begin{equation}\label{eq:B.8}
    \dot{\sigma}(t) = \mathcal{W} \sigma(t) + \sigma(t) \mathcal{W}^{T} + \mathcal{D}.
\end{equation} 
Its evolution should relax to the stationary point at $\dot\sigma(t)=0$ whenever the system is stable. This ODE system \eqref{eq:B.8} is $n \times n$ dimensional, since $\sigma \in \mathbf{R}^{n \times n}$, hence, the usual stability analysis methods could be employed. Moreover, when the system reaches its steady state~\eqref{eq:B.8} leads to the Lyapunov equation~\eqref{eq:9}.    
}
\section{Numerical Verification of Stability}\label{appC}
\renewcommand{\thesection}{C.\arabic{section}}
\renewcommand{\theequation}{C.\arabic{equation}}\setcounter{equation}{0}
{In this appendix, we compare the stability criterion derived from RHSC, as presented in~\eqref{eq:A.3}, with the results obtained through the numerical integration~\eqref{eq:B.8} at steady state. For this comparison, we use feasible experimentally parameters~\cite{b38}. Specifically, we choose the coupling and feedback parameters $\Lambda = 0.4\kappa$, $\phi=0$, $G_{-} = 0.2\kappa$, and $G_{+} = 0.1\kappa$ that satisfy the stability conditions of the system in the weak coupling limit.}

{Using these selected parameters, the eigenvalues of the drift matrix become $\lambda_1=\lambda_2=(-0.0500 + 0.1323i)\kappa, \lambda_3=\lambda_4=( -0.0228 + 0.0000i)\kappa, \lambda_5=\lambda_6=(-0.8772 + 0.0000i)\kappa, \lambda_7=\lambda_8=(-0.0500 - 0.1323i)\kappa$ while the RHSC in~\eqref{eq:A.3} yields the inequalities: $h_1 = 0.0009\kappa^4 > 0$, $h_2 = 0.0036\kappa^5 > 0$, $h_3 = 0.0027\kappa^6 > 0$. These results confirm that $\textrm{Re}(\lambda_l)<0$ and the conditions of RHSC are satisfied, indicating that the system operates within the stable regime.}

\begin{figure}
    \centering
    \includegraphics[width=\linewidth]{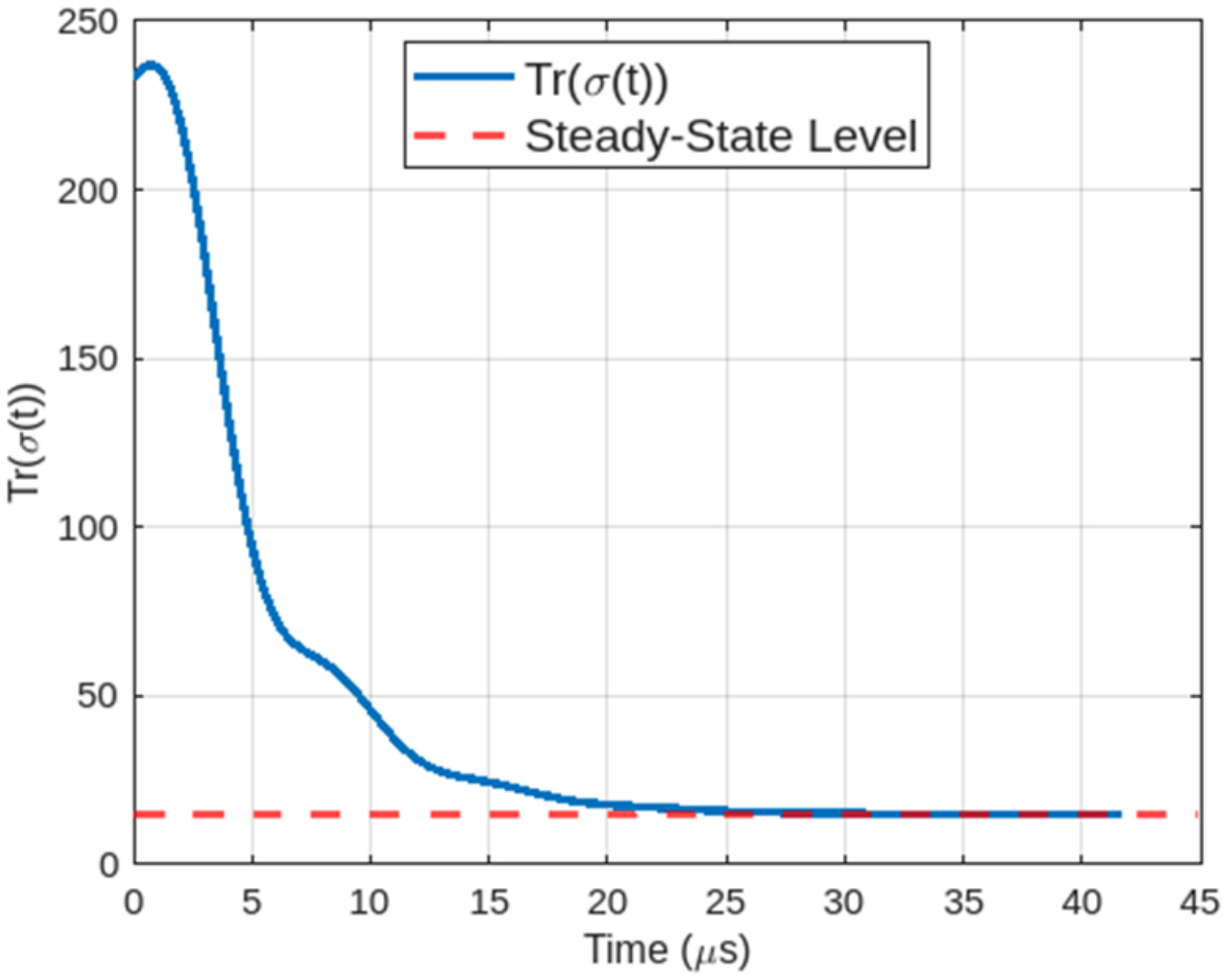}
    \caption{The plot of the trace of the covariance matrix $\textrm{Tr}(\sigma(t))$ versus time for typical parameters $\Lambda = 0.4\kappa$, $\phi=0$, $G_{-} = 0.2\kappa$, and $G_{+} = 0.1\kappa$. The other parameters are the same as Fig.~\ref{fig:2}.}
    \label{fig:9}
\end{figure}

It is also possible to show the stability condition of the system through numerical integration of the time evolution of the CM $\sigma(t)$ in~\eqref{eq:B.8}. As indicated in the main text the matrices $\mathcal{W}$ and $\mathcal{D}$ are constant in time. The initial CM, $\sigma(0)$ is defined on the basis of cavity and mechanical modes initial thermal state. In this regards, we investigate the system's stability under the chosen set of parameters by numerically integrating the differential equation through MATLAB's built-in \texttt{ode45}. The solver \texttt{ode45} is based on a $5^{\text{th}}$-order Runge–Kutta method with a $4^{\text{th}}$-order error estimate for better accuracy. The time evolution of the trace of the CM $\textrm{Tr}(\sigma(t))$ is presented in Fig.~\ref{fig:9}. The trace of the covariance matrix  was tracked and shown to converge smoothly to a steady-state level, confirming the physical stability of the system. Finally, we confirm that the stability results from RHSC (based on eigenvalue analysis of the drift matrix and stability conditions) and those from numerical integration of the ODE are in full agreement. This validates both the mathematical soundness and the physical feasibility of the proposed quantum model.

\section*{Acknowledgments}
This work has been supported by Khalifa University of Science and Technology through the projects C2PS-8474000137 and 8474000739 (RIG-2024-033). The work has been facilitated by the support and resources provided by Adama Science and Technology University. We are grateful to the anonymous referees for their insightful comments and constructive suggestions, which significantly improved the quality and clarity of the manuscript.

\end{document}